\documentstyle [11pt] {article}

\newcommand{\req}[1]{(\ref{#1})}

\newcommand{\kaep}{K\"{a}hler potential}
\newcommand{\kae}{K\"{a}hler}
\newcommand{\yu}{Yukawa couplings}
\newcommand{\wl}{Wilson line}
\newcommand{\wls}{Wilson lines}

\newcommand{\Z}{\ZZ}
\def\bfone{\relax{\rm 1\kern-.35em 1}}
\def\inbar{\vrule height1.5ex width.4pt depth0pt}
\def\IC{\relax\,\hbox{$\inbar\kern-.3em{\mss C}$}}
\def\ID{\relax{\rm I\kern-.18em D}}
\def\IF{\relax{\rm I\kern-.18em F}}
\def\IH{\relax{\rm I\kern-.18em H}}
\def\II{\relax{\rm I\kern-.17em I}}
\def\IN{\relax{\rm I\kern-.18em N}}
\def\IP{\relax{\rm I\kern-.18em P}}
\def\IQ{\relax\,\hbox{$\inbar\kern-.3em{\rm Q}$}}
\def\IR{\relax{\rm I\kern-.18em R}}
\def\ZZ{\relax{\hbox{\mss Z\kern-.42em Z}}}
\font\cmss=cmss10 \font\cmsss=cmss10 at 7pt
\def\ZZ{\relax\ifmmode\mathchoice
{\hbox{\cmss Z\kern-.4em Z}}{\hbox{\cmss Z\kern-.4em Z}}
{\lower.9pt\hbox{\cmsss Z\kern-.4em Z}}
{\lower1.2pt\hbox{\cmsss Z\kern-.4em Z}}\else{\cmss Z\kern-.4em
Z}\fi}

\newcommand{\nwc}{\newcommand}
\nwc{\hyp} {\hyphenation}
\nwc{\be}  {\begin{equation}}
\nwc{\ee}  {\end{equation}}
\nwc{\ba}  {\begin{array}}
\nwc{\ea}  {\end{array}}
\nwc{\bdm} {\begin{displaymath}}
\nwc{\edm} {\end{displaymath}}

\nwc{\bea} {\be\ba{lcl}}
\nwc{\eea} {\ea\ee}

\nwc{\bda} {\bdm\ba{lcl}}
\nwc{\eda} {\ea\edm}

\nwc{\bc}  {\begin{center}}
\nwc{\ec}  {\end{center}}

\nwc{\ds}  {\displaystyle}
\nwc{\bmat}{\left(\ba}
\nwc{\emat}{\ea\right)}
\nwc{\nn} {\nonumber}
\nwc{\nnn} {\nonumber \vspace{.2cm} \\ }
\nwc{\ra}{\rightarrow}
\nwc{\lra}{\longrightarrow}

\nwc{\p} {\partial}
\nwc{\SS} {S}

\nwc{\sieb}{\bf \overline{27}}

\nwc{\scr}  {\scriptstyle}
\nwc{\tx}  {\textstyle}
\nwc{\scs} {\scriptscriptstyle}

\nwc{\ov}  {\overline}

\nwc{\hb}  {\bar h}
\nwc{\xb}  {\bar x}
\nwc{\yb}  {\bar y}
\nwc{\zb}  {\bar z}
\nwc{\wb}  {\bar w}
\nwc{\Ob}  {\bar O}
\nwc{\Yb}  {\bar Y}

\nwc{\ep} {\epsilon}
\nwc{\de} {\delta}
\nwc{\Th} {\Theta}
\nwc{\th} {\theta}
\nwc{\al} {\alpha}
\nwc{\si} {\sigma}
\nwc{\Si} {\Sigma}
\nwc{\om} {\omega}
\nwc{\Om} {\Omega}
\nwc{\Ga} {\Gamma}
\nwc{\ga} {\gamma}
\nwc{\bet} {\beta}
\nwc{\La} {\Lambda}
\nwc{\la} {\lambda}

\nwc{\Sc}  {{\cal S}}
\nwc{\Rc}  {{\cal R}}
\nwc{\Dc}  {{\cal D}}
\nwc{\Oc}  {{\cal O}}
\nwc{\Cc}  {{\cal C}}
\nwc{\gc}  {{\cal g}}

\nwc{\Of}  {{\cal O}_f}
\nwc{\Oft} {{\cal O}_{f_2}}
\nwc{\Ofo} {{\cal O}_{f_1}}

\nwc{\Pc}  {{\cal P}}
\nwc{\Mc}  {{\cal M}}
\nwc{\Ec}  {{\cal E}}
\nwc{\Fc}  {{\cal F}}

\nwc{\Hc}  {{\cal H}}
\nwc{\Kc}  {{\cal K}}
\nwc{\Wc}  {{\cal W}}

\nwc{\Fcp} {{\cal F}^\pr}
\nwc{\Hcp} {{\cal H}^\pr}

\nwc{\Xc}  {{\cal X}}
\nwc{\Gc}  {{\cal G}}
\nwc{\Zc}  {{\cal Z}}
\nwc{\Nc}  {{\cal N}}

\nwc{\xc}  {{\cal x}}
\nwc{\Ac}  {{\cal A}}
\nwc{\Bc}  {{\cal B}}
\nwc{\Uc} {{\cal U}}
\nwc{\Vc} {{\cal V}}
\nwc{\Lc} {{\cal L}}
\nwc{\Qc} {{\cal Q}}

\nwc{\lng} {\langle}
\nwc{\rng} {\rangle}

\nwc{\lf} {\left}
\nwc{\ri} {\right}

\nwc{\diag} {{\rm diag}}
\nwc{\inv}  {{\rm inv}}
\nwc{\mod}  {{\ \rm mod\ }}
\nwc{\dete}  {{\rm det}}
\nwc{\tr}  {{\rm tr}}
\nwc{\im}  {{\rm Im}}
\nwc{\re}  {{\rm Re}}

\nwc{\h} {\frac{1}{2}}
\nwc{\fc} {\frac}

\def\KK{\relax{\rm I\kern-.18em K}}
\def\RR{\relax{\rm I\kern-.18em R}}
\def\NN{\relax{\rm I\kern-.18em N}}
\def\PP{\relax{\rm I\kern-.18em P}}

\def\zz{\relax{\sf Z\kern-.3em Z}}
\def\ZZ{\relax{\sf Z\kern-.4em Z}}
\def\ZZZ{{\relax{\sf Z}\kern -.5em Z}}
\def\ZZZ{Z\kern -0.37em Z}
\def\QQ{{\rm \kern .25em
             \vrule height1.4ex depth-.12ex width.06em\kern-.31em Q}}
\def\CC{{\rm \kern .25em
             \vrule height1.4ex depth-.12ex width.06em\kern-.31em C}}
\if@twoside\oddsidemargin25mm
\else\oddsidemargin25mm\evensidemargin25mm\marginparwidth25mm\fi%
\begin{document}

\begin{flushright}TUM--HEP--213/94\\
hep-th/9412196\\
September 1994
\end{flushright}

%proc
%\baselineskip11pt

\bc
{\large  \bf Low--Energy Properties of (0,2) Compactifications$^\ast$}
\ec

\vskip .5cm

\bc
{\large Peter Mayr\ \  {\small and}
{\ \large Stephan Stieberger}}\\[5mm]
{\large \em Institut f\"{u}r Theoretische Physik} \\
{\large \em Physik Department} \\
{\large \em Technische Universit\"at M\"unchen} \\
{\large \em D--85747 Garching, FRG}
\ec
\begin{quote}
\vskip 5mm \vskip0.5cm
%\hrule width 5.cm \vskip 1.mm
{\small\small
\noindent $^\ast$
Invited talk given by S. Stieberger at the 28th International Symposium on
the Theory of Elementary Particles in Wendisch--Rietz, August 30--September
3, 1994.
Supported by the Deutsche Forschungsgemeinschaft and the EU under contract
no. SC1--CT92--0789.\\
%$^{\ast \ast}$ Adress after January, 1: CERN, Theory
%Division, CH--1211 Geneva, Switzerland.
}
\normalsize
\end{quote}
%\bc
%{\bf Abstract}
%\ec
\begin{quote}
\hrule width 16.cm \vskip 1.mm
{\bf Abstract.} We investigate the low--energy properties of
a $\Z_{12}$ orbifold
with continuous Wilson line moduli. They give rise to a (0,2) superstring
compactification. Their K\"ahler potentials and Yukawa couplings are
calculated. We study the discrete symmetries of the model and their
implications on the threshold corrections to the gauge couplings as well
as for string unification.
\end{quote}

String theory is the only known theory which consistently unifies all
interactions. To make contact with the observable world
one constructs the field--theoretical low--energy limit
of a given ten--dimensional string theory. One possibility to get a
four--dimensional effective N=1 supergravity theory is to compactify six
of the ten dimensions on an internal
Calabi--Yau manifold (CYM) \cite{chsw} or its singular limits,
the toroidal orbifolds \cite{dhvw}.
In general CYMs are Ricci--flat \kae\  manifolds.
If the spin connection is identified with the gauge connection
the gauge group is always $E_6 \times E_8$. There are alternative
embeddings of the spin connection involving stable, irreducible,
holomorphic $SU(4)$ or $SU(5)$ bundles which result in the gauge groups
$SO(10) \times E_8$ or $SU(5) \times E_8$, respectively \cite{w2}.
CYMs with $E_6$ gauge
group\footnote{We will drop the second $E_8$ factor since it couples only
gravitionally to the $E_6$ and therefore plays the r\^ole of the hidden
sector gauge
group giving rise to gaugino condensation.} have matter representation in the
{\bf 27}, $\sieb$ of $E_6$. In addition there can be many singlets. Those
of them which are in the adjoint representation {\bf 8} of the $SU(3)$
are related to the bundle of endomorphisms End T, where T is the holomorphic
tangent bundle over the CYM $K$.
For the standard embedding the number of the ${\bf 27}$ and $\sieb$ generations
is given by topological invariants, the number
of independent harmonic (2,1) forms and (1,1) forms on $K$, respectively.
On the other hand the number of $E_6$ singlets from the {\bf 8} is the
dimension\footnote{${\rm H}^1{\rm (End T)}$ is the space of infinitesimal
deformations of the complex structure of T. Its dimension depends on the
complex structure of
$K$ and therefore on the (2,1)--moduli.}
of ${\rm H}^1{\rm (End T)}$. Of course,
to make contact with the Standard Model one
would like to obtain the gauge group $SU(3) \times SU(2) \times U(1)$.
One way to achieve this is to give Planck--scale vacuum expectation values to
certain components of {\bf 27} or $\sieb$  matter fields. This is an explicit
symmetry breaking lifting the flat directions in the superpotential
\cite{w2,finq}.
On the CYM this can be understood as a deformation of the bundle
that describes the embedding of the spin connection into the gauge group
to a new, stable bundle.
In general this leads to a (0,2) superconformal symmetry on the
world--sheet.
If the CYM is not simply
connected
the group can also be broken by non-trivial holonomies of
gauge fields in the internal directions such that  the vacuum state gauge field
$A_i^a$ cannot necessarily be gauged away, even though the field strength
$F^a_{\ ij}$ vanishes everywhere in order that supersymmetry (SUSY)
remains unbroken \cite{w1,w2}. Therefore the Wilson loop

\be
U(\gamma)=\Pc\ e^{-i \oint\limits_{\gamma} T^a A^a_i dx^i} \ ,
\ee
with $\gamma$ being a closed path on $K$ and $T^a$ the $E_6$ group
generators,  can represent an element of $E_6$ different from unity.
Let us denote the subgroup generated by the elements
$
%\footnote{A CYM defined by the quotient $K=K_0/G$ has
%the fundamental group $\pi_1(K)=G$ and $\Hc=G$, where $G$ is a discrete
%group acting freely on $K_0$. If the symmetry is not freely acting, there may
%be additional massless states in the twisted sectors as e.g. in orbifold
%compactifications. The fundamental group is then smaller than $G$
%%\cite{dhvw}.}
$
$U(\gamma)$ for a fixed
vacuum state
gauge field $A_i^a$ by $\Hc$. At energies below the compactification scale,
$E_6$ will be
spontaneously broken to the subgroup $G$ of $E_6$ which commutes with $\Hc$.
This symmetry breaking is due to the effective Higgs vacuum expectation
values (vevs) of the order of the compactification scale

\be
H^{IJ}=\Pc exp\lf[ -i \oint\limits_\ga (T^a)^{IJ} A_i^a dx^i\ri]
\ee
in the adjoint
%${\bf 248}$ representation of $E_8$ or
{\bf 78} representation of $E_6$.
If the CYM has an Abelian fundamental group, independent Wilson loops have to
commute and therefore $U(\gamma)$ takes the form

\be
U_i=e^{i \sum\limits_{I=1}^{rk G_0} a_i^I H^I},
\ee
where the $H^I$ are generators of the Cartan subalgebra of the original
unbroken gauge
group $G_0=E_6,E_8$ and the real
parameters $a_i^I$ are the {\em Wilson line moduli} corresponding to a breaking
direction in the root space. Toroidal orbifolds have six independent
non--contractible loops which give rise to six \wls\  $a_i^I$ \cite{hpn1,hpn2}.
Since a group generator of the unbroken gauge
group specified by its root $\al$ has to commute with $\Hc$, the roots of
the
unbroken gauge group have to fulfill the condition:

\be
\sum_{I=1}^{rk G_0} \al_I a_i^I=0\mod 2\pi\ ,\ i=1,\ldots,6.
\ee
Note that the above sum is proportional to the mass of the vector boson
corresponding to $\al$ after an adjoint symmetry breaking. The various {\bf 27}
and $\sieb$ representations of
$E_6$ split into representations w.r.t. the unbroken gauge group.
In models with \wls\ the usual relations arising from the organization of
states in $E_6$ multiplets are less stringent since only the singlets
of the combined action of the holonomy and gauge transformations
survive as massless states ~\cite{w1}.
This gives rise  to an
elegant solution of the doublet--triplet splitting problem usually present in
GUTs and even in (2,2)--string models. Moreover also the usual Yukawa
coupling unification of GUTs is absent. On the other hand
gauge coupling unification is still present
since the gauge bosons come from the single adjoint {\bf 78} representation of
$E_6$.
Heterotic string compactifications on a CYM represent a (2,2) superconformal
field theory (SCFT) on the
world--sheet with central charges $(c,\bar c)=(6,9)$ together with free fields
(in light
cone gauge the remaining free left--handed gauge fermions,
one complex left--moving and right--moving boson and one complex right--moving
fermion).
Since the Wilson lines only couple to the free fields describing the unbroken
gauge
degrees of freedom, the right--handed N=2 SCA is completely unaffected by them.
On the other hand the left--handed N=2 algebra can be broken by a
non--standard embedding, but not by the Wilson lines \cite{aadf}.
Anyway this (0,2) SCFT is enough to ensure N=1 space--time supersymmetry
under certain additional conditions \cite{bd}.

The model we want to study is a $\Z_{12}$ orbifold with the torus lattice
$\La_6=SU(3) \times F_4$. In the complex basis the twist has the eigenvalues
$\th=\exp[1/12(4,1,-5)]$. The twist embedding in the gauge lattice $\La_{16}
=E_8 \times E_8'$ is chosen to be $\Th= \Z_3^{(2)} \times \Z_{12}^{(6)}
\times \Z_3'^{(2)} \times \Z_2'^{(2)}$ \cite{holo}.
The resulting gauge
group is $U(1)^2 \times SU(3)^3\times SU(4)'^2 \times
U(1)'^2$. The continuous Wilson lines in the first $E_8$ are chosen to be:

\be
a_1^I=(\la,\mu;0,0;0,0;0,0)\ \ \ ;\ \ \ a_2^I=(-\mu,\la-\mu;0,0;0,0;0,0)\ \ \
{\rm
with}\ \ \ \la,\mu \in \RR\ ,
\ee
w.r.t to the two weights $d_1=(1/\sqrt 2,1/\sqrt 6 ),
d_2=(-1/\sqrt 2,1/\sqrt 6)$ of $SU(3)$.
The corresponding gauge group is

\be
\ba{lcl}
U(1)^2 \times SU(3)^3\times SU(4)'^2 \times U(1)'^2&,& \la,
\mu \in \ZZ \wedge \la+\mu \in 3\ZZ\ ,\\
U(1)^8\times  SU(4)'^2 \times U(1)'^2&,&\la,\mu \in \ZZ \wedge \la+\mu
\notin 3\ZZ\ ,\\
U(1)^6\times SU(4)'^2 \times U(1)'^2&,& \la,\mu \notin \Z\ .
\ea
\ee
At special values of $\la,\mu$ the gauge group is enhanced. Away from these
points
the gauge fields become massive. This fact is known as the stringy
Higgs effect.
Their mass is governed by the \wl\ moduli and the compactification radius $R$

\be
M_X^2\sim \fc{\la^2-\la \mu +\mu^2}{R^2}.
\ee

A N=1 supergravity up to second order derivatives in space--time is completely
characterized by
three functions: the \kae\  potential, the superpotential $W$ and the
so--called $f$--function.
The kinetic terms for the massless fields  are encoded in the \kae\  potential.
The superpotential contains the Yukawa couplings as well the gauge kinetic
terms specified by the $f$--function. The latter determines
the tree--level gauge coupling $g^{-2}_a=\re f_a$.
Our aim is to determine these three functions for some (0,2) orbifold
compactifications. In particular we are interested in their moduli dependence.
Specifically they will depend on the six--dimensional moduli $T$ and $U$
as well as on the complex Wilson line modulus $\cal A$ (containing $\mu$
and $\lambda$).
Therefore we will concentrate on the kinetic terms
$K_{i\bar j}=\fc{\p^2 K(\Phi^a,\bar \Phi^a)}{\p \Phi^i \bar \p \Phi^j}$
for ($\Phi^a \in T,U,{\cal A}$)
and on the corresponding dependence of the Yukawa couplings. The superpotential
reads

\be
W=h_{abc} A_a A_b A_c+ Y_{ijk}(T,U,{\cal A}) \si_i \si_j \si_k+ h'_{ija} \si_i
\si_j A_a\ .
\ee
Here $h_{abc}$ are the \yu\ between three untwisted string states $A_a$ which
are constant. The coupling $Y_{ijk}(T,U,{\cal A})$ between three strings
$\si_i$ from
the
twisted sector is moduli--dependent
due to world--sheet instantons \cite{dfms}.
The last coupling between two strings from the twisted sector and one from
the untwisted sector is in general moduli--dependent as well \cite{dfms,ejss}.
One should however keep in mind that this simple cubic renormalizable form
is corrected after integrating out the heavy string states. These corrections
are non--polynomial in the charged fields.
The cubic, renormalizable superfield couplings dictated by
$E_6$ group theory are $d_{ijk} \Si^i \Si^j \Si^k,\ d^{abc} \Om_a \Om_b \Om_c,
\phi^3$ and $\phi \Si^i \Om_i$ with $\Si^i$ representing a ${\bf 27}$
and
$\Om_a$ a $\sieb$, respectively. The last coupling is important for neutrino
masses as well
as for symmetry breaking if e.g. a $\Si^i$ field acquires a vev of the
order of $M_{\rm Planck}$ \cite{w2}. After the breaking of $E_6$ those fields
which are
not invariant under the twist and the \wl\ action must be set to zero.
The scalar $\phi$ in the $\phi \Si^i \Om_i$ coupling cannot be any
moduli field: this coupling  would give rise to mass terms for the ${\bf 27}$
generations
and thus violate the relation between topology and the number of
generations\footnote{There is another way to see this: $\phi$ being a
(1,1)--modulus would violate the Peccei--Quinn symmetry which is supposed to
hold at least pertubatively. On the other hand there can be couplings as $U
\phi \phi$ with $U$ being a $(2,1)$--modulus \cite{w2}.
An example for a non-renormalizable coupling is $f(T,U)\Si^i \Om_i$, which
generates a $\mu$--term \cite{agnt2}.}.
The superpotential does not get any corrections from
sigma model
perturbation theory as well as from string loop corrections. The only
potential
correction to it comes from world--sheet instantons and
non--perturbative string effects. From instanton corrections
there can arise UV--divergent
$({\bf 27} \sieb)^K$ ($K\geq2$) terms which destabilize the vacuum due to a
non--vanishing beta--function.
The conditions for which such couplings are
absent have been determined  in \cite{distler}.

To get the moduli-- and \wl\ dependence of the \kae\  potential and
the Yukawa
couplings one has to start with a supersymmetric non--linear sigma model rather
than performing the calculation in field theory. We consider the
non--linear sigma--model action \cite{nsw,grv}:

\be
\hspace{-.5cm}S=\int_{C^2} dz d \bar z \sum_{i,j=1}^{6} \bar \p X^i
(g_{ij}+b_{ij}+\fc{1}{4}a_i^{\ I} a_{Ij})\p X^j
+\sum_{I=1}^{16} \sum_{i=1}^{6} \bar \p X^i a_{iI} \p X^I +\sum_{I,J=1}^{16}
\bar \p X^I (G_{IJ}+B_{IJ})\p X^J
\label{action}
\ee
with the chirality constraint $\bar \p X^I=0, P_R^I=0$. It
contains only those gauge fields which could
possibly break the gauge group via the Wilson line mechanism.
As explained before they carry an index of the Cartan
subalgebra. The remaining matter fields can be neglected. These fields can be
easily included
in our results along the lines of \cite{clo}. The complete action
with
all matter fields contains in addition also a world--sheet sigma--model
anomaly.
After discarding the \wls\ independent part \req{action}
can be rewritten
in complex coordinates $Z^i=X^{(2i-1)}+U X^{(2i)},\bar Z^{\bar i}=X^{(2i-1)}+
\bar U X^{(2i)}\ ,\ i=1,2$ with $U=-\h+\fc{i}{2} \sqrt{3}$
being the fixed
$(2,1)$--modulus:

\be
S= \fc{-i}{2\sqrt 3} \int_{C^2} dz d \bar z \
(\tilde T \bar \p \bar Z^{\bar 1}
\p Z^1+\Ac \bar \p \bar Z^{\bar 1} \p Z^2+
T_{E_8^{\perp}} \bar \p \bar Z^2 \p Z^2+{\rm hc.})
\ee
with $T_{E_8^{\perp}}=\fc{2}{3}i\sqrt 3$ and
\bea
\tilde T&=&2 b+i\sqrt 3 (R^2+ \fc{1}{8}  |{\cal A}|^2) \ , \nnn
\Ac&=& \mu+i \fc{\sqrt 3}{3} (2\la-\mu)\ .
\eea
The action and the moduli--space are identical to those of
a four--dimensional $\ZZ_3$ orbifold \cite{clo} at the special points
$T_{12}=0, T_{21}=\Ac, T_1 \equiv \tilde T, T_2 \equiv T_{E_8^{\perp}}$,
where the moduli--space
collapses from a $\fc{SU(2,2)}{SU(2)\times SU(2)\times U(1)}$ coset to
a $\fc{SU(2,1)}{U(1)}$. Therefore the most general \kaep\ being consistent
with the symmetries of that moduli space \cite{ms4} can be
written\footnote{This was recently
shown by a different method in
\cite{clm}.}:

\be
K=-\ln[(-i \tilde T+i \bar{\tilde T})(-i T_{E_8^{\perp}}+i \bar
T_{E_8^{\perp}}) -|\Ac|^2]\ .
\ee
Similar one derives the \kaep\ of the other orbifolds \cite{ms5,sthesis}.

In the following we want to evaluate the Yukawa coupling between twisted matter
fields:

\be
Y_{\Fc_i,\Fc_j,\Fc_k} \equiv
\lim_{|x| \ra \infty} |x|^{4h}
\lng \si^+_{\Fc_i}(x,\bar x)\si^+_{\Fc_j}(1,1)
\si_{\Fc_k}^{--}(0,0)\rng\ .
\label{yu}
\ee
Here $\si^+_{\Fc}(x,\bar x)$ denotes a twist field with conformal weight $h$
corresponding
to the fixed point
$\Fc=(f,F)$
satisfying $\th f=f+2\pi w\ ,\Th F=F+2 \pi W$ and
$\Th \tilde F=\tilde F+2\pi W+2\pi w^ia_i$ with
$w \in \La_6, W \in
\La_{16}$. It is defined via
its operator product
expansion with the coordinate differentials $\p Z(z,\bar z)$ and
creates a twisted string at the world--sheet
insertion $z=x$,
where the local monodromy becomes \cite{dfms}:

\be
\ba{llcccrcr}
\ds{X^i(e^{2\pi i}z,e^{-2\pi i}\bar z)} &=& \ds{(\theta^i_{\ j} X^j)
(z,\bar z)}\ , \nnn
\ds{X^I(e^{2\pi i}z,e^{-2\pi i}\bar z)} &=& \ds{(\Theta^I_{\ J} X^J)
(z,\bar z)\ .}
\ea
\label{bouncond}
\ee
One observes that the local monodromy conditions do not feel the \wl\
whereas the global
monodromy conditions become \wl\ dependent \cite{sthesis}. The number of
fixed points $\tilde F$ is not changed in the presence of
continuous \wls\ \cite{hpn2}. The fixed points are subject to the fixed--point
selection
rule $\Fc_i+\Fc_j-(1+\Th)\Fc_k \in \Lambda $. Using results of \cite{ejss} one
arrives
at \cite{ms5,sthesis}:

\be
Y_{\Fc_i,\Fc_j,\Fc_k}(\tilde T,\Ac) \sim (\det g)^{\fc{1}{4}} \sum_{\vec v \in
(1-\th^2)(f_k-f_j+w),w \in \La_6
\atop \vec u
\in (1-\Th^2)(F_k-F_j+\La_{16}+w^ia_i)^\perp}
e^{i\pi (T |v|^2+\Ac \bar v u+T_{E_8^{\perp}} |u|^2)}\ .
\label{sss}
\ee
Again, we discarded the \wl\ independent parts which can be
obtained from \cite{ejss}.
$u,v$ are the complexified
components $v=v^1+iv^2$ and $u=u^1+iu^2$ of $\vec v,\vec u$, respectively.
One observes that the \wls\ produce additional
hierarchies between the \yu.

Let us now turn to the topic of threshold corrections to the gauge couplings.
It can be shown that the expression of \cite{vk} can be simplified in
the N=2 sector to \cite{ms5,sthesis}:

\be
\triangle_a(\tilde T,\bar{\tilde T},\Ac,\bar \Ac)=b_a^{(1,\Th^3)} (\Ac=0)
\int\limits_{\tilde \Ga} \fc{d^2 \tau}{\tau_2}
\sum_{k_1,k_2 \in \ZZZ}Z^{4d}(\tau,\bar \tau,\tilde T,\bar{\tilde T},\Ac,\bar
\Ac,k_1,k_2)
\Cc_a(\tau,k_1,k_2)\ .
\label{res0}
\ee
with
%($T=2b+i\sqrt 3 R^2$)\ ,
\be
%\ba{rcl}
Z^{4d}=\sum_{n^1,n^2 \in \ZZZ \atop m_1,m_2 \in \ZZZ}e^{\fc{-\pi
\tau_2}{\im(\tilde T-i \sqrt 3/8 |\Ac|^2)\im U}|\tilde T U n^2+\tilde T
n^1-U(m_1+\Ac k_1-\h \Ac k_2)+m_2-\h \Ac k_1|^2}e^{2 \pi i\tau(m_1n^1+m_2n^2)}
%\ea
\ee
and a holomorphic moduli--independent function $\Cc_a(\tau,k_1,k_2)$.
$b_a^{(1,\Th^3)}(\Ac=0)$
is the $\beta$--function coefficient of the $(1,\Th^3)$ sector for $\Ac=0$.
The integrand
is invariant under $\Ga_0(2)$ as it is required by modular invariance.
The evaluation of \req{res0} is rather cumbersome and is the subject of
\cite{ms5,sthesis}. Instead let us discuss the discrete symmetries of
$\triangle_a(\tilde T,\bar{\tilde T},\Ac,\bar \Ac)$. One finds the
following symmetries \cite{ms4}:

\bea \label{a1}
\tilde T &\lra&\ds{-\fc{1}{\tilde T}\ ,\ \Ac \lra  \fc{\Ac}{\tilde T}}\nnn
\tilde T&\lra&\tilde T+1\nnn
\Ac&\lra&\ds{\Ac+2U\ ,\ \tilde T\lra \tilde T+U-\la \bar U-\mu U}\nnn
\Ac&\lra&\ds{\Ac+2\ ,\ \tilde T\lra \tilde T+U-\la-\mu \bar U}
\eea
In addition one has $\Ac \ra \Ac-2\mu-2\la U$ together with the corresponding
transformation on $\tilde T$.
There are also some fixed directions in the $(\la,\mu)$ or $\Ac$ plane, along
them a shift in $\Ac$ is not accompanied by a shift in the field $b$. E.g.:
along $\mu=0$ or $\la=0$ the shifts $\la\ra\la+6$ or $\mu\ra\mu+6$
respectively, lead to the same theory.
Eqs. \req{a1} are also the symmetries of the N=2 spectrum. We want
to stress that the
symmetries of the \kaep\ are in general not the symmetries of the
threshold corrections since the truncation performed in these cases violates
modular invariance. Therefore one has to be careful in
identifying automorphic functions of $SU(2,1)$ and threshold corrections. For
more details see \cite{sthesis}.

Finally we want to discuss the implications of \req{res0} for gauge coupling
unification in string theory. At string tree--level all gauge couplings are
related to the gravitional coupling by the well--known equations

\be
g_a^2k_a = 4\pi \alpha^{\prime -1}G_N = g^2_{\rm string}\ , \; \;
\forall\;  a \ ,
\ee
valid at $M_{\rm string}$. Here $\alpha^{\prime}$ is the  inverse
string tension and $k_a$ is the
Kac--Moody level of the group factor labeled by $a$. The scale
$M_{\rm string}$ can be  determined to be $M_{\rm string} = 0.52 \, g_{\rm
string} \times 10^{18}\, {\rm GeV}$ in the
$\overline{\rm DR}$ scheme \cite{vk}. Taking $g_{\rm
string} \sim 0.7$ corresponding to the GUT--coupling constant one obtains
a $M_{\rm string}$ which is one order of
magnitude larger than the unification scale $M_{\rm X} \sim 2 \cdot
10^{16} {\rm GeV}$ obtained in  the minimal supersymmetric Standard Model
(MSSM). On the other hand threshold corrections due to the infinite many string
states with
masses above $M_{\rm string}$ can split the one--loop gauge couplings at
$M_{\rm string}$.
This splitting could allow for an effective unification at a scale $M_{\rm X} <
M_{\rm string}$ or destroy the unification. The Georgi, Quinn and Weinberg
equations
for the evolution of the gauge couplings below $M_{\rm string}$ read \cite{vk}:

\be \label{run}
\fc{1}{g^2_a(\mu)}=\fc{k_a}{g^2_{\rm string}}+\fc{b_a}{16 \pi^2} \ln
\fc{M_{\rm
string}^2}{\mu^2}
+\fc{1}{16\pi^2}\triangle_a \ .
\ee
These equations also allow us to determine $\sin^2 \th_{\rm W}$ and
$\al_{\rm S}$ at $M_{\rm Z}$. However to obtain phenomenological viable values
one needs huge threshold corrections because of the large value of $M_{\rm
string}$ at tree--level \cite{IL}.
The  threshold corrections can be divided into a field dependent and a constant
part. The gauge group dependent part of the constant piece has been
calculated in
\cite{vk} for some (2,2) and in \cite{mns} for (2,0) orbifold
compactifications. It turns out that these corrections are surprisingly mild
and cannot remove the discrepancy between $M_{\rm X}$ and $M_{\rm string}$.
Remarkably these threshold corrections are proportional to the
beta--function coefficients and therefore only shift the unification scale
while preserving the unification scenario. The gauge group dependent
part\footnote{Recently, the universal part has been discussed in \cite{kk}.}
of field--dependent threshold corrections has been calculated in
\cite{vk,DKL2,ant2,ms1}.
In particular moduli dependent contribution can give rise to threshold
corrections of the deliberate size for appropriate vevs of the background
fields. However it was shown in \cite{IL} that assuming only a
$(T,U)$--dependence one needs vevs which are unnatural far away from the
self--dual points. A simple argument based on the symmetries (\ref{a1}) shows
that the Wilson line dependence of the thresholds is comparable to that on the
$T$
and $U$ fields thus offering the interesting possibility  of large thresholds
with less exceptional background configurations \cite{ms4,ms5,sthesis}.
\rm\baselineskip=14pt

\ \\
{\bf Acknowledgements.} We would like to thank Hans Peter Nilles for helpful
discussions. St. St. is also grateful in particular to Harald
Dorn, Dieter L\"ust and Gerhard Weigt for organizing the meeting and
providing a pleasant atmosphere
at Wendisch--Rietz.

%proc
%\vskip2cm

\small \small
\ \\
{\bf REFERENCES}
%\newpage

%NEW MACRO FOR BIBLIOGRAPHY
\newcommand{\bibit}{\it}
\newcommand{\bibbf}{\bf}
\renewenvironment{thebibliography}[1]
        {\begin{list}{[\arabic{enumi}]}
        {\usecounter{enumi}\setlength{\parsep}{0pt}
%1.25cm IS STRICTLY FOR PROCSLA.TEX ONLY
\setlength{\leftmargin 0.5cm}{\rightmargin 0pt}
%0.52cm IS FOR NEW DATA FILES
%\setlength{\leftmargin .52cm}{\rightmargin 0pt}
         \setlength{\itemsep}{0pt} \settowidth
        {\labelwidth}{#1.}\sloppy}}{\end{list}}


\begin{thebibliography}{9}
%\begin{thebibliography}{99}

%\begin{thebibliography}{99}




\newcommand{\np}{\mbox{\em {Nucl. Phys.} {\bf B }}}
\newcommand{\pl}{\mbox{\em {Phys. Lett.} {\bf B }}}
\newcommand{\cmp}{\mbox{\em {Comm. Math. Phys. }}}
\newcommand{\prd}{\mbox{\em {Phys. Rev.} {\bf D }}}
\newcommand{\pr}{\mbox{\em {Phys. Rep.\ }}}
\newcommand{\mpl}{\mbox{\em Mod. Phys. Lett.\ } {\bf A}}
\newcommand{\bi}[1]{\bibitem{#1}}



\bi{chsw} P. Candelas, G. Horowitz, A. Strominger and E. Witten,
       \np {\bf 258} (1985) 46

\bi{dhvw} L. Dixon, J. Harvey, C. Vafa and E. Witten,
       \np {\bf 261} (1985) 678; {\bf B 274} (1986) 285;\\
       L. E. Ib\'a\~nez, J. Mas, H. P. Nilles and F. Quevedo,
       \np {\bf 301} (1988) 157

\bi{w2} E. Witten, \np {\bf 268} (1986) 79

\bi{finq} A. Font, L. E. Ib\'a\~nez, H. P. Nilles and F. Quevedo,
      \np {\bf 307} (1988) 109

\bi{w1} E. Witten, \np {\bf 258} (1985) 75

\bi{hpn1} L. E. Ib\'a\~nez, H. P. Nilles and F. Quevedo,
\pl {\bf 187} (1987) 25;\\
L. E. Ib\'a\~nez, J. E. Kim, H. P. Nilles and F. Quevedo,
\pl {\bf 191} (1987) 283

\bi{hpn2} L. E. Ib\'a\~nez, H. P. Nilles and F. Quevedo,
\pl {\bf 192} (1987) 332

\bi{aadf} G. Athanasiu, J. Atick, M. Dine and W. Fischler,
      \pl {\bf 214} (1988) 55

\bi{bd} T. Banks and L. Dixon, \np {\bf 307} (1988) 93

\bi{holo} T.J. Hollowood and R.G. Myhill, {\em Int. J. Mod. Phys.} {\bf A3}
(1998) 899

\bi{dfms} L. Dixon, D. Friedan, E. Martinec and S. Shenker,
      \np {\bf 282} (1987) 13

\bi{ejss} J. Erler, D. Jungnickel, M. Spali\'{n}ski and S. Stieberger,
      \np {\bf 397} (1993) 379

\bi{agnt2} I. Antoniadis, E. Gava, K.S. Narain and T.R. Taylor,
     \np {\bf 432} (1994) 187

\bi{distler} J. Distler, \pl {\bf 188} (1987) 431; J. Distler and B. Greene,
\np {\bf 304} (1988 ) 1

\bi{nsw} K. S. Narain, M. H. Sarmadi and E. Witten, \np {\bf 279} (1987) 369

\bi{grv} A. Giveon, E. Rabinovici and G. Veneziano, \np {\bf 322} (1989) 167

\bi{clo} M. Cveti\v{c}, J. Molera and B. Ovrut, \prd {\bf 40} (1989) 1140;\\
M. Cveti\v{c}, J. Louis and B. Ovrut, \pl {\bf 206} (1998) 227


\bi{ms4} P. Mayr and S. Stieberger, {\em Discrete symmetries and threshold
corrections in (0,2) orbifold compactifications},
Munich preprint TUM--HEP--208/94 to appear

\bi{clm} G. Lopes Cardoso, D. L\"{u}st and T. Mohaupt, \np {\bf 432} (1994) 68

\bi{ms5} P. Mayr and S. Stieberger, {\em K\"{a}hler potentials,
Yukawa couplings and threshold corrections to gauge couplings in (0,2)
orbifold compactifications}, Munich preprint TUM--HEP--212/94 to appear

\bi{sthesis} S. Stieberger, {\em One--Loop corrections and string unification
in (2,2) and (0,2) superstring compactifications}, PhD thesis at the Technische
Universit\"at M\"unchen, Munich preprint to appear

\bi{vk} V. Kaplunovsky, \np{\bf 307} (1988) 145 and Erratum, Stanford
ITP--838/1992

\bi{IL} L. E. Ib\'a\~nez and D. L\"{u}st, \np {\bf 382} (1992) 305

\bi{mns} P. Mayr, H. P. Nilles and S. Stieberger, \pl {\bf 317} (1993) 53

\bi{DKL2} L. Dixon, V. Kaplunovsky and J. Louis \np {\bf 355} (1991) 649

\bi{ant2} I. Antoniadis, E. Gava and K.S. Narain,  \pl {\bf 283} (1992) 209

\bi{ms1} P. Mayr and S. Stieberger, \np {\bf 407} (1993) 425; \np {\bf 412}
(1994) 502

\bi{kk} E. Kiritsis and C. Kounnas, {\em Curved four--dimensional spacetime
as infrared regulator in superstring theories}, CERN--TH.7471/94
(hep--th/9410212)


\end{thebibliography}
\end{document}